\documentstyle[prl,aps,epsf,multicol]{revtex}
\begin{document}
\draft

\def\i{\imath\,}
\def\ih{\frac{\imath}{2}\,}
\def\undertext#1{\vtop{\hbox{#1}\kern 1pt \hrule}}
\def\ra{\rightarrow}
\def\lfa{\leftarrow}
\def\ua{\uparrow}
\def\da{\downarrow}
\def\Ra{\Rightarrow}
\def\lra{\longrightarrow}
\def\ler{\leftrightarrow}
\def\lrb#1{\left(#1\right)}
\def\O#1{O\left(#1\right)}
\def\EV#1{\left\langle#1\right\rangle}
\def\tr{\hbox{tr}\,}
\def\trb#1{\tr\lrb{#1}}
\def\dd#1{\frac{d}{d#1}}
\def\dbyd#1#2{\frac{d#1}{d#2}}
\def\pp#1{\frac{\partial}{\partial#1}}
\def\pbyp#1#2{\frac{\partial#1}{\partial#2}} 
\def\pd#1{\partial_{#1}}
\def\br{\\ \nonumber & &}
\def\brr{\right. \\ \nonumber & &\left.}
\def\inv#1{\frac{1}{#1}}
\def\be{\begin{equation}}
\def\ee{\end{equation}}
\def\bea{\begin{eqnarray}}
\def\eea{\end{eqnarray}}
\def\ct#1{\cite{#1}}
\def\rf#1{(\ref{#1})}
\def\EXP#1{\exp\left(#1\right)} 
\def\INT#1#2{\int_{#1}^{#2}} 
\def\LHS{left-hand side }
\def\RHS{right-hand side }
\def\COM#1#2{\left\lbrack #1\,,\,#2\right\rbrack}
\def\AC#1#2{\left\lbrace #1\,,\,#2\right\rbrace}

\title{Electron fractionalization and cuprate superconductivity}
\author{T. Senthil}
\address{Massachusetts Institute of Technology,
77 Massachusetts Ave.,  Cambridge 02139, USA.
}

\date{\today}
\maketitle

\begin{abstract}
We discuss the possibility that the electron may be fractionalized in some quantum phases of matter in two or higher dimensions.
We review the theory of such phases, and show that their effective theory is a $Z_2$ gauge theory. 
These phases may be characterized theoretically through the notion of topological order. We discuss
the appeal of fractionalization ideas for building theories of the cuprates, and possible ways of testing these ideas.

\end{abstract}

\vspace{0.15cm}


\begin{multicols}{2}
\narrowtext

\section{Introduction}
\label{intro}
A system of strongly interacting electrons can have several unusual properties. 
Among the most remarkable is the possibility that the electron itself breaks apart
into several more ``fundamental'' constituents in such a system. More precisely, the 
excitations of the many electron ground state may have quantum numbers that are fractions of those of the 
electron. 
The possibility of obtaining fractional quantum numbers for excitations in a 
solid is an old theme in condensed matter physics. 
There are two famous well-studied examples of the phenomena. The first involves electronic systems 
in one spatial dimension\cite{1deg} as realized by polymers such as polyacetylene or, more recently, by 
Carbon nanotubes. In such a situation, the electron decays into several other excitations under rather
general conditions. The second example comes from two dimensional electron gases in strong magnetic fields that show 
the fractional quantum Hall effect\cite{QHEbook}. It is 
basic to the theory of the effect that there be excitations that carry a fraction of the charge of the electron, 
which have indeed been observed in experiments\cite{QHEexp}.

In the last several years, there has been a growing suspicion amongst some workers in the field that 
fractionalization of the electron could occur under conditions that are far less restrictive than the two examples 
mentioned above. In particular, it appears possible that two or three dimensional systems in weak or zero magnetic fields, could already 
display quantum phases with fractional quantum numbers. The physics behind this
fractionalization however seems to be different from the much better understood one dimensional example.

Much of the interest in understanding the theoretical possibility of fractionalization
comes from the high temperature superconducting materials. There are a large number of phenomena displayed 
by these materials (including the superconductivity itself) which have appealing simple explanations
in terms of the idea that the electron is fractionalized. Consequently, fractionalization ideas
have played a central role in several different attempts\cite{PWA,KRS,LN,topth} to build theories of the cuprates. However, a complete
understanding of the cuprates is still lacking.

In contrast to the cuprates, where the suspicion of fractionalization is 
fueled mainly by very suggestive experiments, there are a growing number of 
other systems where theoretical calculations suggest the possibility of fractionalized electrons. Numerical calculations
of the properties of the solid phase of electrons in two spatial dimensions\cite{2dwcr,lhull} suggest a range of densities where the
system may have fractionalized excitations. Similar results have also been obtained for the closely related system of the solid phase of 
$He-3$ in two dimensions\cite{lhull}. There have also been a number of theoretical 
studies of Heisenberg spin models on exotic lattices\cite{marst1,marst2} as realized in some
recent experiments\cite{shsuth,coldea} which have been argued to be promising candidates to look for fractionalization. 
Taken together, the experiments on the cuprates and the frustrated magnets\cite{coldea}, 
and the theoretical work on these other systems seem to suggest that the 
phenomenon of fractionalization of the electron is perhaps much more common than may have been imagined fifteen years ago.

In the first part of this paper, we will discuss the theory of  fractionalized phases in spatial dimensions 
greater than one, and in zero external magnetic fields. 
There has been considerable progress\cite{RSSpN,Wen1,NLII,z2long,topth,qdm1} 
in understanding the properties that such phases must possess, if they do exist. 
A number of different approaches have been pursued to theoretically access fractionalized phases. Interestingly, 
these different approaches all give the same final description. 
The second part of the paper will consider the 
application of fractionalization ideas to the high temperature superconductors.

\section{Theoretical description of fractionalized phases}
\label{tdofp}
We begin with a discussion of the theoretical description of fractionalized phases. For concreteness, we also 
specialize to two spatial dimensions. What do we mean by a ``theoretical description'' of any phase of matter? Let us imagine a
microorganism living inside a material deep in the given phase. For the time being, we restrict attention to quantum phases, and we will
imagine that the system is at zero temperature. What will the universe be as seen by this living being?  
In particular, we may ask about the ``elementary particles'' inside this universe and their interactions.

Consider, for instance, a conventional solid in a band insulating phase. In the universe inside such a solid, there are
elementary excitations with the quantum numbers of the electron, and there is a finite non-zero energy gap to 
create them. These excitations interact with each other through the Coulomb interaction. Other ``particles'' which are
composites made by binding some integer number of these excitations are also in principle possible. 

Now consider an insulating solid in a fractionalized phase. For conceretness, we will focus on one example of such a 
phase (which is possibly relevant to the high-$T_c$ materials). In this phase, there are three distinct ``elementary particles''. 
First, there is a particle with charge $e$ (equal to the electron charge)
and spin $0$ which has bosonic statistics. We will refer to this as the ``chargon''. Second, there is a particle with charge $0$
and spin $1/2$ with fermionic statistics. We will refer to this as the ``spinon''. Finally, there is a strange third ``vortex-like'' particle
which has no charge and no spin - we will refer to this as the ``vison'' for reasons that will become clear later.  Note that none of these particles 
have the quantum numbers of the electron. To obtain excitations with the quantum numbers of the electron, it is necessary to bind
together a chargon and a spinon. Clearly, the term ``fractionalized'' is an appropriate description of a phase with this 
structure of elementary excitations. 

As we are specifically discussing insulating phases, the chargon is a gapped excitation. The spinon, on the other hand, may or may not be gapped.
In contrast, as we will see later, it is necessary for the vison to be gapped. But what are the attributes of the vison if it does not carry any charge or spin?
To answer this question, it is useful to first ask about the interactions between these elementary particles in this fractionalized universe. 
As the chargons carry electrical charge, they interact with each other through the Coulomb interaction. 
The spinons would also interact with each other and with the chargons through various short-ranged interactions
(allowed by the symmetries of this universe). But the most crucial interaction is a long-ranged one between the vison and the other two particles. 
If a chargon (or spinon) is taken around a loop that encircles the vison, the wavefunction of the system changes sign. In other words, the 
chargon (or spinon) sees the vison as a source of $\pi$ flux. This is the principle attribute of the vison. If the loop
encloses two visons, there is no effect on the wavefunction of the system when a chargon or spinon is transported once around. 
Thus, two visons is equivalent to having no visons at all; in other words, visons can be created or annihilated in pairs. Thus the 
visons only have a conserved $Z_2$ quantum number. 

The frustration of the chargon or spinon motion by the visons is what is at the heart of the requirement that they be gapped in the 
fractionalized phase. Indeed, if the visons could themselves move freely in the ground state, the chargons and spinons would not be able to propagate
as independant particles. Note that an electron suffers no phase change when it encircles a vison. Thus, if visons proliferate in the ground state, 
it becomes impossible for the fractions of the electron to propagate unless they are confined together to form an electron. Thus, the gapping of
the visons is necessary for the fractions of the electron to have any integrity at all in the first place. 

The long-range interaction between the vison and the two fractions of the electron may be given a concise mathematical description as follows: 
As already noted, we may think of this interaction as due to an Aharonov-Bohm phase induced by the vison that is seen by the 
chargons and spinons. To capture this, we may assign a $Z_2$ charge each to the spinon and the chargon, and a $Z_2$ gauge flux to the vison. 
This will then encapsulate the long-range interaction between the various excitations. The ``fundamental'' theory of this fractionalized universe
will then take the form of a $Z_2$ gauge theory. What is a Hamiltonian of this theory? To answer this, it is first useful to
assemble together the information we have on the symmetries of this universe. This will constrain the structure of the Hamiltonian.
Imagine that the system lives on a square lattice, and define operators $b_r, f_{r\alpha}$ on the sites $r$ such that
$b^{\dagger}_r$ creates a chargon at site $r$ while $f^{\dagger}_{r\alpha}$
creates a spinon with spin $\alpha = \ua, \da$ at site $r$. The Hamiltonian should clearly respect the 
following symmetries:
\itemize
\item
{\em Electrical charge conservation}:

This implies that the Hamiltonian should be invariant under a global phase change of the chargon 
operators:
\be
b_r \ra b_r e^{i\theta}.
\ee

\item
{\em Spin conservation}:

This implies that the Hamiltonian be invariant under a global spin rotation
\be
f_{r\alpha} \ra U_{\alpha \beta}f_{r\beta}.
\ee
Here $U$ is an arbitrary constant $SU(2)$ matrix.
Note that the spinon number $f^{\dagger}_rf_r$ does {\em not} need to be conserved by the dynamics. This important observation
will have interesting physical consequences, as we will see later. 

\item
{\em $Z_2$ gauge symmetry}:

As discussed above, a concise mathematical encapsulation of the long-range interaction between the 
chargons (or spinons) and the visons is provided by assigning a $Z_2$ gauge charge to each of the
former and regarding the vison as the corresponding flux. As is usual with gauge interactions, it is much more convenient
to formulate the theory in terms of a gauge field rather than in terms of the gauge flux itself. We therefore
introduce operators $\sigma^z_{rr'}$ that live on the bonds of the lattice, and are to be thought of as 
$Z_2$ gauge fields. The flux of this gauge field through any elementary plaquette is given by the product of the 
$\sigma^z_{rr'}$ on the four bonds of the plaquette. 
The
Hamiltonian must then be  invariant under the $Z_2$ gauge transformation $b_r \ra -b_r, 
f_r \ra -f_r$ at any site $r$ of the lattice accompanied by letting $\sigma^z_{rr'}
 \ra -\sigma^z_{rr'}$ on all the links connected to that site. 

We may now construct a Hamiltonian subject to the constraints imposed by these symmetries. 
\bea
\label{z2gh}
H & = & H_c + H_{\sigma} + H_s, \\
H_c & = & -\sum_{<rr'>} t_c \sigma^z_{rr'} \left(b^{\dagger}_r b_{r'}
+ h.c \right) + U \sum_r \left( N_r - 1 \right)^2, \\
H_s & = & -\sum_{<rr'>} \sigma^z_{rr'}\left[t_s \left(f^{\dagger}_r f_{r'} + 
h.c \right) \right. \nonumber \\
+ & & \Delta_{rr'} \left. \left(f_{r\ua} f_{r'\da} - f_{r \da} f_{r' \ua}  + h.c \right) \right], \\
H_{\sigma} & = & -K\sum_{\Box} \prod_{\Box} \sigma^z_{rr'} - h \sum_{<rr'>} \sigma^x_{rr'}.
\eea   
The first term $H_c$ describes the physics of the chargons. The
operator $N_r = b^{\dagger}_rb_r$ measures the number of bosons at site $r$. For simplicity, we have
specialized to integer filling, {\em i.e}, to an average of one charge per site. The constant $t_c$ 
are the chargon hopping amplitudes and the constant $U$ describes a local chargon repulsion. In principle, we should 
also include the long-range Coulomb interaction between the chargons. The assumption that the system
is insulating implies that $U >> t_c$. 

The second term $H_s$ describes the physics of the spinons. The first term in $H_s$ describes hopping of the spinons on the lattice
with some hopping amplitude $t_s$. The second term is also quadratic in the spinon operators but allows for the 
creation (or destruction) of a singlet pair of spinons. Note that such pairing terms are forbidden 
for {\em electrons} in an insulator as they would violate charge conservation symmetry. 
In contrast, the spinon Hamiltonian only needs to conserve the total physical spin. Thus, singlet ``pairing'' terms
such as the term multiplied by $\Delta_{rr'}$ are allowed in the spinon Hamiltonian. Here 
the constant $\Delta_{rr'}$
contains the information about the ``pairing'' symmetry of the spinons. Thus, the separation of spin
and charge that occurs in the fractionalized insulators under consideration generically allow pairing terms for the 
neutral spinon excitations. 
This is an important point whose physical consequences we shall examine later. 
The details of the spin physics will depend on the orbital symmetry of the $\Delta_{rr'}$. A particularly interesting 
situation to consider in the context of the cuprates  is a $\Delta_{rr'}$ that has $d_{x^2 - y^2}$ symmetry.

\begin{figure}
\epsfxsize=3.5in
\centerline{\epsffile{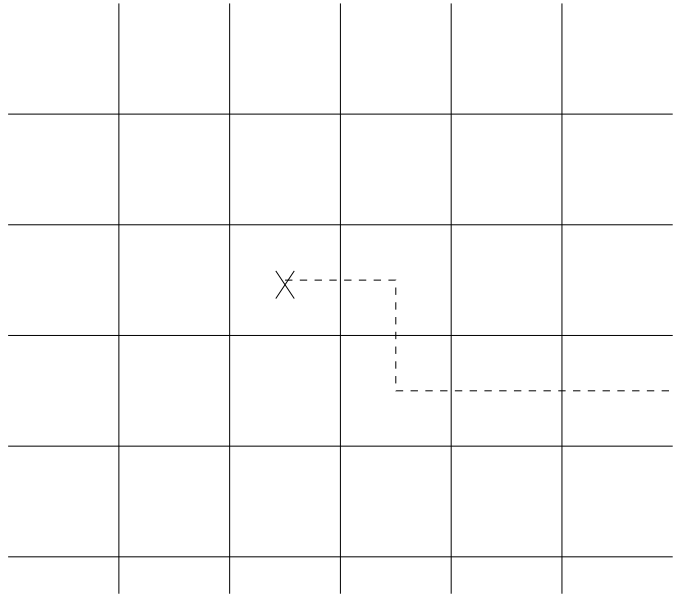}}
\vspace{0.15in}
\caption{Vison state at $h = 0$. The field $\sigma^z_{rr'} = -1$ on all the bonds cut by the dashed line. On all other bonds,
$\sigma^z_{rr'} = 1$.
}
\vspace{0.15in}
\label{vison}
\end{figure}

The third term $H_{\sigma}$ represents the physics of the gauge fields, that is, the dynamics of the vison.  
The operator $\sigma^z_{rr'}$ was introduced above as the $Z_2$ gauge field. The box product refers to a product over 
the four bonds that make up an elementary plaquette. The operator $\sigma^x_{rr'}$ 
is the corresponding $x$ component of the Pauli spin operator. The presence of the term proportional to $h$ gives 
dynamics to the visons. To see this, and to get some intuition for this piece of the Hamiltonian, consider first the limit 
$h = 0$. In that limit, we may fix the value of each $\sigma^z_{rr'}$. The ground state of $H_{\sigma}$ in this limit is 
simply one in which there is no flux through any plaquette (we assume that $K > 0$), {\em i.e}, $\prod_{\Box} \sigma^z_{rr'} = 1$
for every plaquette. An excited state with one vison may be constructed by allowing for a flux of $-1$ through a single 
plaquette. This will require changing the sign of a string of bonds that intersect a line drawn from the center of the 
plaquette to infinity (see Fig. \ref{vison}). This single vison state will cost an energy $2K$. 
Now consider introducing a small but non-zero $h$. The values of $\sigma^z_{rr'}$ at each bond may no longer be fixed. 
The $h$ term is readily seen to allow the motion of the flux from one plaquette to a neighbouring plaquette. Thus it allows the vison
move. As $h$ increases, the vison gap will decrease. At some value of $h$, we may expect that the vison gap will
collapse to zero. As the Hamiltonian above is specifically constructed to describe a phase with gapped visons, it is necessary 
to assume that $K$ is sufficiently bigger than $h$ so that the vison gap is non-zero.

In the Hamiltonian above, we have only written down the simplest terms consistent with the symmetries that are present in the fractionalized
phase. In principle, other more complicated terms that are allowed by the symmetries could also be added. These include four spinon interaction terms
that could, for instance, induce magnetic ordering or other short range interactions between the chargons and the spinons.      

This Hamiltonian must be supplemented with  
the constraint equation
\be
\label{constr}
G_r = \Pi_{r' \in r}\sigma^x_{rr'} e^{i\pi \left(f^\dagger_r f_r + N_r\right)} = 1.
\ee
Here the product over $\sigma^x_{rr'}$ is over all links that emanate
from site $r$.
The operator $G_r$, which commutes with the
full Hamiltonian, is the generator of the local $Z_2$ gauge symmetry. Thus the 
constraint $G_r =1$
simply expresses the condition that the physical states in the Hilbert space are those that are gauge
invariant. 

This concludes our discussion of the theoretical description of fractionalized phases. We have taken the point of view of a
microorganism living inside such a fractionalized material. Such a being would discover the presence of the chargon, spinon, 
and the vison as ``elementary particles'' in it's universe, and the symmetries governing their interactions. A concise mathematical 
description of these interactions would be provided by a $Z_2$ gauge theory Hamiltonian with the structure described above.

\section{Theoretical approaches to fractionalization}
\label{tatf}
The previous section has probably left the reader feeling somewhat uncomfortable. We seemed to have described the 
final structure of the fractionalized phase without indicating how one may arrive at such a description
by starting with models of interacting electrons. The purpose of doing so was to separate the physical
properties of the fractionalized phase from properties of the theoretical techniques that may be used
to access such a phase when starting with models of interacting electrons. Indeed, one may obtain fractionalization
from different starting points in terms of electrons and following very different routes. In this section, we will
describe some of these routes.

\subsection{Gauge theories}
\label{gt}
The field of strongly interacting electron systems is replete with various kinds of gauge theories. Superficially, 
distinct gauge theories with
gauge group $U(1), SU(2)$ and $Z_2$ have all appeared\cite{u1,LN,su2,z2long} in the literature in the context of theories of the cuprates. 
It is extremely important, however, to realize that there actually are two very different invocations of theories with
gauge structure. In one, the gauge structure is a mathematical way of encapsulating the physical interactions between the 
true fractionalized excitations of the phase. In this case, the gauge theory description of the phase is a necessity.
This is the point of view adopted above, in Section \ref{tdofp}.
The other completely different appearance of gauge theories arises in attempts to calculate 
properties of specific models of strongly interacting electrons when it is often convenient to change variables from 
electron operators to other degrees of freedom that formally create objects that carry fractional quantum numbers. 
This formal change of variables is often associated with some redundancy which then expresses itself as a gauge symmetry
when the theory is recast in terms of the new variables. At this stage, this change of variables and the related gauge 
structure are simply statements about the method of calculation rather than about the physics of any given phase. 
In particular, even conventional phases of matter (where the excitations are {\em not} fractionalized) may, in 
principle,  be described in this basis. Clearly, this kind of gauge symmetry should not be confused with the other kind described in 
Section \ref{tdofp}
- the latter is a physical statement about the interactions between the physical excitations in a fractionalized phase.

As a concrete example, consider the square lattice nearest neighbour antiferromagnetic Heisenberg spin model defined by the
Hamiltonian
\be
\label{heis}
H = J \sum_{<rr'>} \vec S_r . \vec S_{r'}
\ee
Here the $S_{r}$ are spin-$1/2$ operators, and $J > 0$ so that the interaction is antiferromagnetic. 
It is known that the ground state of the Hamiltonian has long ranged Neel order. Furthermore, there is strong evidence 
that the system is {\em not} in a fractionalized phase. The excitations in this phase are magnons which will have a gapless 
linear dispersion at low energies. 
 
{\em Formally}, we may  represent the 
spin operators as bilinears of boson operators:
\be
\vec S_r = \frac{1}{2}s^{\dagger}_r \vec \sigma s_{r}
\ee
where the $s_r = (s_{r\ua}, s_{r\da})$ is a two-component boson. 
This is a faithful representation of the spin operator so long as we impose the constraint 
\be
s^{\dagger}_rs_r = 1
\ee
An identical representation using fermionic rather than bosonic
operators is also possible. This change of variables introduces a gauge redundancy: we may let
$s_{r\alpha} \ra e^{i\theta_r}s_{r\alpha}$ at each site $r$ with $\theta_r$ an arbitrary angle at each site. Clearly, the physical
spin operators are left unchanged under this transformation. Consequently, the theory when reexpressed in terms of the $s_{r\alpha}$
will have a local $U(1)$ gauge symmetry. 

The model described above is in a conventional Neel ordered phase. In principle, this may be described
in terms of the $s_{r\alpha}$ and the associated $U(1)$ gauge symmetry.  
Do the $s_{r\alpha}$ fields or the $U(1)$ gauge symmetry have any direct physical meaning in this 
phase? It seems that they do not. 

\begin{figure}
\epsfxsize=3.5in
\centerline{\epsffile{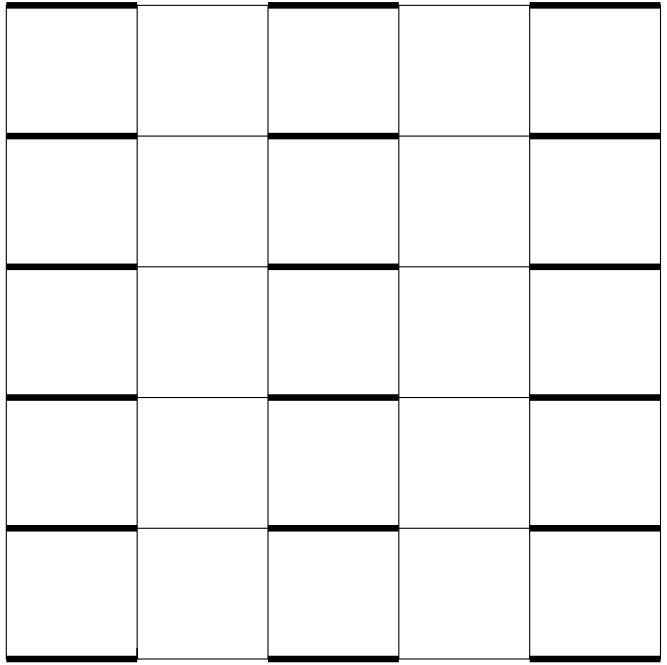}}
\vspace{0.15in}
\caption{Caricature of the columnar spin-Peierls state. The thick bonds denote the locations of the singlet bonds.
}
\vspace{0.15in}
\label{dimsol}
\end{figure}

Now consider modifying this model by adding frustrating further neighbour or other kinds of interactions. 
In principle, a wide variety of phases then become possible depending on the precise nature of the additional interactions. 
These will include several phases with no magnetic long range order. These phases may be further characterized by 
asking for whether they display excitations with fractional spin quantum numbers or not. An example of a phase with
no magnetic long range order or fractionalization of spin is the columnar spin-Peierls phase depicted in Fig. \ref{dimsol}.
In this phase, each spin forms a singlet bond with a nearest neighbour, and the singlet bonds arrange themselves in columns. 
The lowest energy spin-carrying excitation is a triplet of {\em gapped} magnons. This phase too is in principle contained in the 
model when reformulated in terms of the $s_{r\alpha}$ and the associated gauge symmetry. However, once again neither the 
$s_{r\alpha}$ or the $U(1)$ gauge field are physical excitations of this phase. This is true even if the change of variables to the 
$s$ operators is a useful trick to perform approximate analytic calculations on the original model.

We now turn to a phase of these spin models that does display excitations with fractional quantum numbers. Such  phases
were described in Ref. \cite{RSSpN,Wen1} and the excitations are a gapped spin-$1/2$ spinon, and a vison. (Note that there are no charge
degrees of freedom in these spin models). The spinon acquires a phase 
of $\pi$ on encircling the vison. Thus, an effective field theory for the excitations in this phase takes the form of a 
$Z_2$ gauge theory.

Accessing a fractionalized phase in the $s$-representation requires condensing a singlet pair\cite{RSSpN,Wen1,MudFr} formed out the $s$-operators - such a singlet
pair carries spin $0$ and, obviously, physical electric charge $0$. However, under the local $U(1)$ gauge transformation, it carries charge $2$.
Consequently, a singlet pair condensate ``breaks'' the gauge symmetry down to $Z_2$ from $U(1)$. 
The physical spinon excitations are the analogs of the BCS
quasiparticles of the singlet pair condensate. Note that the $U(1)$ gauge charge of the 
bare $s$ operator is screened out by the condensate.
However as the singlet pairs carry gauge charge $2$, there still is a remnant $Z_2$ gauge charge. 
Naively, one might have expected (by analogy to a BCS superconductor) that the singlet pair 
condensate will also support vortex excitations which carry 
flux of the $U(1)$ gauge field that is quantized in integer units of $\pi$. However, due to subtle lattice effects, 
this gauge flux is only well-defined modulo $2\pi$. Specifically, monopole configurations at which a gauge flux of an integer
multiple of $2\pi$ can apear or disappear are allowed in the theory. 
Consequently, the vortices only carry a $Z_2$ quantum number. They are the visons. 

All the phases mentioned above are obviously contained in the formulation with a local $U(1)$ symmetry in terms of the $s$ operators.
However, this gauge structure is completely unrelated to that describing the physical excitations of the phase. In the conventional
phases with no fractionalization of spin, there is no ``statistical''  long ranged interaction between the excitations, and consequently no
need for a gauge description of the physics. In the fractionalized phases, there is a long-ranged ``statistical'' interaction that is 
quite adequately described as a $Z_2$ gauge interaction.

The discussion above considered fractionalization in frustrated spin models. 
In the phases described in Ref. \cite{RSSpN}, the spinons had bosonic statistics. 
However, as pointed out in Ref. \cite{ReCh}, it is possible to change the statistics
of the spinons (or the chargons) by binding to a vison. Thus the issue of statistics 
depends on energetics determined by the details of the model Hamiltonian. 
(See Ref. \cite{gof} for more discussion of this issue).

If we consider microscopic models with charge fluctuations, 
it is possible to directly derive a Hamiltonian of the form Eqn. \ref{z2gh}, but in the limit $K << h$. 
In particular, a general class of interacting electron models with strong spin and pairing fluctuations 
can be recast\cite{z2long} in terms of ``chargon'' operators $b_r$ and ``spinon'' operators $f_{r\alpha}$. This change
of variables introduces a degree of redundancy in the description. Specifically, all physical observables are invariant under a local 
change in the sign of the spinon and chargon operators. This implies that the resultant theory must have a local $Z_2$ gauge invariance. 
At this stage, this representation is morally on the same footing as the representation of spin models in terms of the $s$ operators
as discussed above.

The actual values of the various parameters
such as $t_c$ or $K$ or $h$ are of course dependent on the details of the system. Upon coarse graining the theory by integrating out high
energy fluctuations, the values of these parameters would change. The lower energy properties would be 
described by a {\em renormalized} Hamiltonian of the form in Eqn. \ref{z2gh}  with altered parameters\cite{note1} which are non-trivial functions of the original parameters. 
The properties of the system 
would be determined by this new Hamiltonian. In view of this, we 
could imagine examining the properties of the Hamiltonian in Eqn. \ref{z2gh} for {\em arbitrary} values of the parameters. It can be easily argued that
in the limit $K >> h$, there is a fractionalized phase where the 
physical excitations in such a phase are created by renormalized versions of the chargon and spinon operators. In addition, the vison will
directly be described as the vortex excitation of the $Z_2$ gauge field that is necessarily present in this formulation. This phase will have a finite 
region of stability in the space of parameter values. Consequently, it is reasonable to expect that for some choices of the 
original {microscopic} electron model, the system will end up in a fractionalized phase that is described as in Section \ref{tdofp}.

Another way of understanding the connection between Hamiltonians of the form Eqn. \ref{z2gh}, and models written in terms of electron variables
is to explicitly consider the limit where  $h$ is larger than all the other couplings. 
In this limit, the field $\sigma^x_{rr'}$ is forced by the large $h$ term to predominantly
have the value $+1$:
\be
\sigma^x_{rr'} \approx 1
\ee
This implies that the $\sigma^z_{rr'}$ must have large fluctuations. Indeed, the $\sigma^z_{rr'}$
causes virtual fluctuations out of the $\sigma^x_{rr'} = +1$ 
space into states with $\sigma^x_{rr'} = -1$. These can be treated in second order perturbation theory leading to an effective 
Hamiltonian in terms of the $b$ and $f$ alone. Consider, in particular, the extreme limit $K = 0$, and $U = \infty$. In this limit, the 
chargon number is fixed to be one per site 
\be
N_r = 1.
\ee
In this limit, there are no charge fluctuations. The constraint equation \ref{constr} then simply becomes
\be
(-1)^{f^{\dagger}_rf_r} = -1.
\ee 
Thus, the constraint 
restricts the number of spinons to be {\em odd} at each site. 
This is of course equivalent to requiring that there be exactly one spinon at each site:
\be
f^{\dagger}_rf_r = 1.
\ee 
Thus, the Hilbert space in this limit, is exactly the same as in a spin model. Furthermore, with $K =0$, 
the $\sigma^z_{rr'}$ at different bonds are
decoupled from each other and can be integrated out independent of each other. Performing the perturbation theory to second
order, we will get some interaction term between the spins at nearest neighbour sites. Clearly, this interaction term will
need to be spin rotation invariant, and the only such possibility is a simple Heisenberg exchange. The sign may easily be shown to be antiferromagnetic.
We thus end up with the Hamiltonian Eqn. \ref{heis} with $J$ determined by $t_s, \Delta$ and $h$. 

If we move away from the special limit above, the procedure of integrating out the gauge field will generate some electron Hamiltonian
with short ranged hopping and various kinds of short ranged interactions at least for $K << h$. Thus, the gauge theory Hamiltonian 
is indeed a rewrite of legitimate electronic models with short ranged interactions.

We may now argue as above for the existence and stability of fractionalized phases in {\em some} electron models
with short ranged interactions.  
What is lost in this approach (or the other ones discussed below)
is the precise connection between the original microscopic model and the parameter regime of the gauge Hamiltonian which admits 
a stable fractionalized phase. Consequently, we are unable as yet to produce a specific microscopic electron model which is unambiguously in a fractionalized phase.

\subsection{Vortex pairing}
\label{vp}
An alternate approach\cite{NLII,z2long} that provides further insight to accessing fractionalized insulating phases comes from viewing it in terms of the excitations 
of a superconductor. For a system of interacting bosons in two spatial dimensions at a density commensurate with an underlying lattice, 
there are two possible phases - a superfluid phase and a bosonic Mott insulator. It is well known that it is possible to set up
a dual representation\cite{xydual} in which the physics is formulated in terms of vortices in the boson wavefunction rather than the particles themselves. 
In this dual approach, the superfluid state is described as a state in which the vortices are gapped. The insulator, on the other hand, is described
as a condensate of vortices. Thus, the dual description provides a view of the insulator in terms of the excitations of the superfluid.
For the electronic systems under consideration here, we again start from a BCS superconducting phase. A useful observation, due to Kivelson and Rokhsar\cite{KR}, 
is that, in some loose sense, a (singlet) superconductor 
already has separation of spin and charge. If one imagines inserting an electron into
the bulk of a superconductor, its charge gets screened out by the condensate
to leave behind a neutral spin-carrying excitation - a ``spinon''. To mathematically implement this idea, consider the action of a BCS superconductor
(assumed to be $s$-wave for simplicity) coupled to phase fluctutations:
\bea
S & = & S_{\varphi} + S_{qp} \\
S_{\varphi} & = & \int d^3 x \frac{\kappa_{\mu}}{2}\left(\partial_{\mu} \varphi \right)^2 \\
S_{qp} & = & S_{kin} + S_{\Delta} \\
S_{kin} & = & \int d^2 x d\tau \sum_{\alpha} c^{\dagger}_{\alpha} \left(\partial_{\tau}- \frac{\nabla^2}{2m} - E_f \right)c_{\alpha} \\
S_{\Delta} & = & \int d^2 x d\tau \Delta 
\left(e^{-i\varphi}c_{\ua}c_{\da} + c.c \right)
\eea
Here $\varphi$ is the Cooper pair phase, and $c^{\dagger}_{\alpha}$ is the creation operator for an electron with spin $\alpha$. The
strong coupling between the Cooper pair phase and the electrons in the last term is inconvenient to handle. It is 
therefore convenient to change variables to neutral objects by absorbing {\em half} the phase of the Cooper pair field into the electron operator.  
\be
c_{\alpha} = e^{i\frac{\varphi}{2}}f_{\alpha}
\ee
The $f$ operators formally create neutral spin-$1/2$ fermionic excitations which we may identify with spinons. The theory can now, in principle, be 
reformulated in terms of these objects and the phase field $\varphi$. The pairing term in the action simply becomes
\be
S_{\Delta} = \int d^2x d\tau \Delta \left(f_{\ua}f_{\da} + c.c \right)
\ee
However, this change of variables is not single-valued in the 
presence of vortex configurations in the phase field\cite{NLII}. Around a $hc/2e$ vortex, the Cooper pair phase winds by $2\pi$. 
As the electron operator is single-valued, it
implies that the spinon field $f$ changes sign on encircling a vortex. Thus the spinons ``see'' the $hc/2e$ vortex as a source of $Z_2$ gauge flux.
Now consider destroying the superconductivity at zero temperature through quantum phase fluctuations. If the resulting phase is an insulator, 
this is tantamount to condensing  vortices in the phase field. If however the insulator is obtained by condensation of $hc/2e$ 
vortices, then the spinon motion is severely frustrated and leads to confinement of the spinons. Fractionalization, in this dual view, 
is therefore only possible due to condensation of $hc/e$ vortices\cite{NLII,z2long} 
while the $hc/2e$ ones remain gapped. The spinon motion is {\em not}
frustrated by $hc/e$ vortices, and thus it is possible for the spinons to have integrity in such an insulator. 
What are the other excitations in such a fractionalized phase? Clearly the presence of neutral spin-$1/2$ spinons implies the presence of 
spinless charge $e$ chargons in order to form an electron. This may also be seen more directly as follows. The physical electric charge,
in this dual description of the insulator, is carried by vortices in the vortex condensate. Condensation of $hc/e$ vortices then implies
charge quantization in units of $e$. Thus the chargons are the dual vortices of the paired vortex condensate. 
The vison may also be recovered in this dual description. The $hc/2e$ vortex is gapped and uncondensed in this insulator. However due to the condensation of
$hc/e$ vortices, the vorticity ceases to be a good quantum number. In other words, the vorticity of the $hc/2e$ vortex is screened out by the 
condensate. Thus the $hc/2e$ vortex has only a $Z_2$ quantum number in this phase. In fact, it is the vison.
The vison is thus a $hc/2e$ vortex that has shed it's electromagnetic flux. 

In this view, the long-ranged statistical interaction between the $hc/2e$ vortex and the spinon in the superconductor simply translates into 
one between the vison and the spinon in the fractionalized insulator. Similarly, it can also be shown in this dual description 
that the same statistical interaction also exists
between the vison and the chargon. We thus recover the $Z_2$ gauge structure of the fractionalized phase in this description.

\subsection{Quantum dimer models}
A useful model of a spin-$1/2$ system in a phase with a spin gap is the quantum dimer model\cite{RK}. The model is based on viewing the 
physics of such a system as due to the formation of short ranged singlet valence bonds between pairs of spins. In the limit of
extreme short-range spin correlations, one may hope to restrict attention to valence bond configurations in which each 
spin is paired with a nearest neighbour. Even so, there are a large number of nearest neighbour valence bond configurations
which can fluctuate into each other. The problem of diagonalizing the Hamiltonian in this subspace still needs to be solved. 
A nearest neighbour valence bond can approximately be thought of as a hard-core dimer (if different valence
bonds configurations are approximated to be orthogonal). Thus, the space spanned by the valence bond states corresponds to 
that of the various dimer coverings of the lattice. The original problem is then approximately reduced to solving a quantum 
Hamiltonian for these hard-core dimers that allows for fluctuations between the various dimer coverings. The motivation for, and
the physics of these dimer models is reviewed in much more detail in the recent article by Moessner et. al.\cite{qdm2}

A simple Hamiltonian that governs the physics of these dimers on a square lattice may be written down as follows:

\bea 
&&\lefteqn{H_{QDM} = -t \hat{T} + v \hat{V}=}
\nonumber \\ 
&\sum_\Box&\left\{-t\left
( |
\setlength{\unitlength}{3947sp}%
\begingroup\makeatletter\ifx\SetFigFont\undefined%
\gdef\SetFigFont#1#2#3#4#5{%
  \reset@font\fontsize{#1}{#2pt}%
  \fontfamily{#3}\fontseries{#4}\fontshape{#5}%
  \selectfont}%
\fi\endgroup%
\begin{picture}(155,154)(533,319)
\thicklines
\put(664,343){\circle{18}}
\put(556,342){\line( 1, 0){105}}
\put(557,343){\circle{18}}
\put(557,449){\circle{18}}
\put(556,449){\line( 1, 0){105}}
\put(664,449){\circle{18}}
\end{picture}
\rangle
\langle
\setlength{\unitlength}{3947sp}%
\begingroup\makeatletter\ifx\SetFigFont\undefined%
\gdef\SetFigFont#1#2#3#4#5{%
  \reset@font\fontsize{#1}{#2pt}%
  \fontfamily{#3}\fontseries{#4}\fontshape{#5}%
  \selectfont}%
\fi\endgroup%
\begin{picture}(154,155)(397,321)
\thicklines
\put(527,452){\circle{18}}
\put(528,344){\line( 0, 1){105}}
\put(527,345){\circle{18}}
\put(421,345){\circle{18}}
\put(421,344){\line( 0, 1){105}}
\put(421,452){\circle{18}}
\end{picture}
|+h.c.  \right) +v\left
( |
\setlength{\unitlength}{3947sp}%
\begingroup\makeatletter\ifx\SetFigFont\undefined%
\gdef\SetFigFont#1#2#3#4#5{%
  \reset@font\fontsize{#1}{#2pt}%
  \fontfamily{#3}\fontseries{#4}\fontshape{#5}%
  \selectfont}%
\fi\endgroup%
\begin{picture}(155,154)(533,319)
\thicklines
\put(664,343){\circle{18}}
\put(556,342){\line( 1, 0){105}}
\put(557,343){\circle{18}}
\put(557,449){\circle{18}}
\put(556,449){\line( 1, 0){105}}
\put(664,449){\circle{18}}
\end{picture}
\rangle \langle
\setlength{\unitlength}{3947sp}%
\begingroup\makeatletter\ifx\SetFigFont\undefined%
\gdef\SetFigFont#1#2#3#4#5{%
  \reset@font\fontsize{#1}{#2pt}%
  \fontfamily{#3}\fontseries{#4}\fontshape{#5}%
  \selectfont}%
\fi\endgroup%
\begin{picture}(155,154)(533,319)
\thicklines
\put(664,343){\circle{18}}
\put(556,342){\line( 1, 0){105}}
\put(557,343){\circle{18}}
\put(557,449){\circle{18}}
\put(556,449){\line( 1, 0){105}}
\put(664,449){\circle{18}}
\end{picture}
|+
|
\setlength{\unitlength}{3947sp}%
\begingroup\makeatletter\ifx\SetFigFont\undefined%
\gdef\SetFigFont#1#2#3#4#5{%
  \reset@font\fontsize{#1}{#2pt}%
  \fontfamily{#3}\fontseries{#4}\fontshape{#5}%
  \selectfont}%
\fi\endgroup%
\begin{picture}(154,155)(397,321)
\thicklines
\put(527,452){\circle{18}}
\put(528,344){\line( 0, 1){105}}
\put(527,345){\circle{18}}
\put(421,345){\circle{18}}
\put(421,344){\line( 0, 1){105}}
\put(421,452){\circle{18}}
\end{picture}
\rangle \langle
\setlength{\unitlength}{3947sp}%
\begingroup\makeatletter\ifx\SetFigFont\undefined%
\gdef\SetFigFont#1#2#3#4#5{%
  \reset@font\fontsize{#1}{#2pt}%
  \fontfamily{#3}\fontseries{#4}\fontshape{#5}%
  \selectfont}%
\fi\endgroup%
\begin{picture}(154,155)(397,321)
\thicklines
\put(527,452){\circle{18}}
\put(528,344){\line( 0, 1){105}}
\put(527,345){\circle{18}}
\put(421,345){\circle{18}}
\put(421,344){\line( 0, 1){105}}
\put(421,452){\circle{18}}
\end{picture}
|
\right)\right\}\, ,
\nonumber \\
&&
\label{eq:exqdm}
\eea
Here the first term allows for one dimer configuration to fluctuate to another by locally flipping 
a pair of parallel dimers on the two bonds of a plaquette. 
The second term counts the number of such plaquettes which are capable of being flipped
in any particular dimer configuration. The model is easily generalized to other lattices - see, for instance Ref. \cite{qdm1}.  

In this dimer description of spin gapped phases, there are again two qualitatively distinct phases. 
First there are phases that may be described as a dimer solid. The columnar spin-Peierls state
mentioned briefly in Section \ref{gt} is one such, and simply corresponds to the dimers stacking up in columns as in Fig. \ref{dimsol}.
A dimer solid breaks some of the symmetries of the lattice on which the dimer model is defined. 
The other qualitatively different kind of phase is one in which the dimers do not break any of the lattice symmetries. Such a phase
may be termed a dimer liquid. A prototypical wavefunction for such a state is an equal superposition of all
dimer configurations. 
In terms of the original spin model, such a dimer liquid phase has fractionalized spinon
excitations in it's spectrum (see Ref. \cite{qdm2} for instance). This may be understood as follows: Consider breaking a single valence bond to create two
spin $1/2$ objects. In the dimer langauge, this corresponds to creating two monomers at nearest neighbour sites. 
In a dimer liquid, these monomers may be moved far apart with only a finite energy cost. Thus, one expects deconfined spin-$1/2$ 
excitations to exist in the corresponding spin model.

For the square lattice quantum dimer model above, it has been argued\cite{RSSuN} that the dimer liquid phase is absent. However, 
in a recent advance\cite{qdm1}, a dimer liquid 
phase has been shown to be 
stable on the triangular lattice (see also Ref. \cite{NkScht}). In general, it has been argued that nearest neighbour quantum dimer models on lattices with
non-bipartite structure could possess dimer liquid phases while bipartite lattices do not\cite{SachVoj}.

What are the excitations of a spin gapped system that is well-described as a dimer liquid? As mentioned above, there will
be spin-$1/2$ spinons above the spin gap. Interestingly, there will also be vison excitations in such a system. 
Let the ground state wavefunction of the dimer liquid be 
\be
|gd \rangle = \sum_{C} a_C |C \rangle
\ee
where $C$ refers to a single dimer configuration. In the dimer liquid phase, $a_C$ will have appreciable weight 
over a large number of configurations $C$. In particular, for the proptotypical wavefunction mentioned above, 
$a_C \equiv a$ is independent of $C$. 
To construct a state with a 
single vison on any given plaquette\cite{Kiv}, first introduce a string extending from that plaquette all the way to infinity.
Now construct the wavefunction
\be
\sum_C a'_C |C \rangle
\ee
where $a'_C = a$ for all $C$ where an {\em even} number of dimers cut the string, and $a'_C = -a$ for all $C$ 
where an {\em odd} number of dimers cut the string. 

Now consider including a monomer ({\em i.e} a spinon) 
in a state with a vison. When such a monomer is moved all the way around a loop encircling the vison, it can easily be seen that
the even and odd parts of the wavefunction interchange with each other. Consequently, the total wavefunction
changes sign. This is exactly as expected for a spinon moving around a vison. 

Thus, we recover the spinons and visons with a long range ``statistical'' interaction that is described as a $Z_2$ gauge interaction
in the dimer liquid picture as well. There is actually a fruitful {\em microscopic} mapping\cite{FrKiv,SachVoj,qdm2} between the dimer model 
and various kinds of gauge theories
with $U(1)$ or $Z_2$ symmetry. The gauge structure of these microscopic rewrites of the dimer model of course are not
related directly to the presence or absence of a gauge interaction between the excitations of the phase. In particular, 
in the dimer liquid phase, the physical excitations interact through a $Z_2$ gauge interaction whatever the original microscopic 
representation. Similarly, in the dimer solid that describes a columnar spin-Peierls state, there is no gauge interaction between the physical excitations
of the original spin system. Nevertheless, these microscopic gauge theory representations are useful as {\em calculational
techniques} for analysing the properties of the dimer models. 

The results showing the stability of the dimer liquid phase have taken us a step closer to understanding what 
microscopic models may show fractionalizaion. Nevertheless, it is still not possible to unambiguously point to a microscopic
electron model that shows a fractionalized ground state. In this context, 
some results obtained from the nearest neighbour dimer models
need to be viewed with caution when drawing conclusions about the corresponding spin models. 
For instance, as mentioned above, on the square lattice such a model does not display a dimer liquid phase.
This however is {\em not} to be interpreted as a signature of the impossibility of obtaining fractionalized spin gapped phases, 
even in spin-only models, on the square lattice. The restriction to nearest neighbour dimers is done for convenience of analysis, 
and in the case of the square lattice, introduces a bipartite symmetry which ultimately is resposible for the absence of the dimer liquid phase.
Once further neighbour valence bonds, in particular those that connect points on the same sublattice, are included as additional possible dimers, 
it would presumably become possible for a dimer liquid to be stabilized even on the square lattice. Of course, the representation 
of valence bonds as dimers also needs to be reexamined once these additional dimers are included.

\section{Topological order}
Several conventional phases of matter (for instance a solid or a superfluid) 
may be characterized in terms of the symmetries that are broken spontaneously in that state. 
How do we characterize a fractionalized phase? What, if any, is the fundamental distinction between a fractionalized phase of the 
kind that we have described and one with no fractionalization? 
One may be tempted to say that a fractionalized phase is 
distinguished from a conventional phase by asking for the lowest energy excitation with,
for instance, spin-$1/2$. In the conventional case, this would be an electron which also
carries an electric charge $e$. In the fractionalized phases of the kind discussed 
above, one might expect that the corresponding excitation is a spinon which is charge
neutral. However, this test for fractionalization is fraught with difficulties. Consider
a fractionalized phase where the lowest energy spin-$1/2$ excitation is indeed a spinon. 
Now imagine turning on some attractive interaction between the chargons and spinons
which binds them into an electron at low energies. This could, in principle, happen 
without going through a phase transition. Then, the lowest energy  
excitation with spin-$1/2$ is an electron (as opposed to a spinon) though the 
system is adiabatically connected to a fractionalized phase (see Ref. \cite{NLII}
for a discussion of this effect). Furthermore, other tests such as the vanishing 
of the quasiparticle residue also fail in this situation. 

The precise distinction lies in the presence of the gapped vison excitation in the fractionalized phase. 
The gapping of the visons implies that they are expelled from the interior of a fractionalized sample. 
It is useful to make an analogy with the Meissner effect in a superconductor which may be thought of as the explulsion
of vorticity and the associated electromagnetic flux from the sample. Thus vison expulsion is to fractionalization 
what the Meissner effect is to superconductivity. This observation may be exploited to provide a precise experimental 
characterization of fractionalization\cite{toexp,topth} that may then be used to detect it's presence in various physical systems. 

As discussed in Ref. \cite{ReCh,Wen1}, and elaborated in Ref. \cite{topth}, the expulsion of the 
topological vison excitations leads to a precise topological distinction between fractionalized
and conventional phases. To see this simply, consider a fractionalized insulator of the kind described in Section \ref{tdofp}.
Assume that the system lives on the surface of a cylinder. Now imagine threading a vison through the hole of the cylinder. 
If the vison is gapped in the bulk, it will not be able to escape out of the hole (at zero temperature) in any 
finite time in the thermodynamic limit. Thus there are two topologically distinct sectors for the Hilbert space of the system on the cylinder.
Consider the energy of the ground states in the two sectors. The presence or absence of the vison only affects the boundary conditions 
of the chargons and spinons on encircling a loop that winds once around the cylinder. The boundary conditions are 
periodic in the absence of a vison and antiperiodic with the vison present. 
This change of boundary conditions leads to a 
slight shift of the energy of the chargons and spinons. In the insulating phase under consideration here, the chargon is gapped.
Therefore the change in it's boundary conditions only leads to an energy change 
that is exponentially small in the circumference of the cylinder.
Similarly, if the spinons are gapped, the energy still changes only by an 
exponentially small amount. In a fractionalized phase where the spinons
are paired with $d_{x^2 - y^2}$ symmetry, there will be gapless nodal points. 
Even in the presence of such gapless nodal spinons, the energy change 
due to the change in boundary conditions goes to zero with the cylinder circumference (albeit more slowly as the inverse circumference). 

Thus, we are led to the result that the ground states in the two topological 
sectors have exactly the same energy in the thermodynamic limit. 
This topological ground state degeneracy on non-trivial manifolds provides a precise theoretical distinction 
between a fractionalized and conventional insulator. We emphasize that this degeneracy is over and above any 
degeneracies that may exist in the ground state due to some conventional broken symmetry. Thus, even two states with the same conventional
broken symmetries, may still be distinguished from each other on the basis of whether they are fractionalized or not.

The presence of this topological ground state degeneracy should be extremely useful in numerical calculations on model systems to detect the presence of
fractionalization. Consider, for instance, an exact diagonalization of some two dimensional interacting electron model on a system with periodic boundary conditions for
the electrons. This is equivalent to putting the system on a torus. If the system is in a fractionalized 
insulating phase, there should be four states that become degenerate 
as the system size is increased. 

In his pioneering work on the fractional quantum Hall fluids, Wen\cite{Wen2} introduced the notion of ``topological order'' to 
characterize quantum phases of matter that are topologically distinct from other more conventional phases. Following that terminology, we may 
say that the fractionalized phases discussed here are characterized by the presence of topological order.

\section{High temperature superconductors}
\label{hts}
Perhaps the most important experimental systems driving the theoretical study of 
fractionalized phases in dimensions higher than one are the high temperature superconductors. 
It was suggested early on\cite{PWA,KRS} that the electron was possibly fractionalized in these materials. A considerable 
amount of activity over the last several years have led to elaborate developments of this basic idea. 
Indeed, the idea that the electron is, in some sense, broken apart is a crucial ingredient of 
many different currently popular theories\cite{topth,su2,stripe} of the cuprates. 
We will focus on one particular possibility which builds closely on the theoretical discussion of fractionalization
of the previous sections. 

Superconductivity in the cuprates is obtained by doping ``parent'' compounds 
that 
are Mott insulators -  rendered insulating by  
strong electron-electron interactions. These parent compounds also display 
Neel antiferromagnetism. 
We will assume that the parent Mott insulators are, in addition to their magnetism, fractionalized. This is a strong assumption
and we will discuss suggestive (though not definitive) experimental support for it in Section \ref{ui}. As we discuss below, doping such a fractionalized
Mott insulator naturally leads to superconductivity. Several other qualitative features of the underdoped cuprates, for instance, the 
physics in the pseudogap regime are also natural consequences of doping a fractionalized Mott insulator. We discuss this in Section \ref{pg}. 

\subsection{Fractionalization as a route to superconductivity}
\label{frts}
The cuprate high-$T_c$ materials are amongst the most complicated 
systems studied extensively in solid state physics. 
In addition to the high temperature superconductivity itself,
they display a wide variety of novel phenomena.
It is hoped by many that underlying this remarkably 
complex behaviour, might lie a simple explanation
which will give insight into the mechanism of
the superconductivity. The idea that the electron is fractionalized in these materials
indeed provides an elegant and simple explanation of the superconductivity
and other properties.  Remarkably, fractionalization
provides a
novel route
to superconductivity which dispenses entirely with
the notion of electron pairing.
Quite generally, to obtain superconductivity in a many-body system it is necessary
to condense a charged particle. In an electronic system
the naive route would be to 
condense the electron, but this is of course
not possible as the electron is a fermion. 
The BCS solution was to argue that a weak  
attractive interaction between the electrons (or more precisely
between Landau quasiparticles) binds them into pairs,
which condense as a charge $2e$ boson.
But fractionalization describes an altogether different
{\it route} to superconductivity.
Once the electron is fractionalized, it's charge is no longer tied to it's Fermi statistics. 
The resulting charged boson (the chargon) can then directly condense
leading to superconductivity.

Note that the chargon is gapped in the Mott insulator. Upon doping, the charge density is no longer commensurate with the underlying crystal
lattice. At extremely small doping, Coulomb interaction effects presumably dominate leading to an insulating phase with charge ordering.
As the doping is increased, the chargon kinetic energy dominates leading to superconductivity. 

As an immediate consequence, it is clear that 
the superfluid density $\rho_s$ of this superconductor will be quite small and would decrease with decreasing doping. Specifically,
it will be set by the deviation of the chargon density from that in the Mott insulator, {\em i.e} the doping $x$ rather than $1 - x$ as 
is expected in a mean field BCS superconductor derived from a Fermi liquid. Thus, we naturally obtain 
$\rho_s \propto x$ as is well-established empirically in the underdoped cuprates. 
It should be emphasized that this particular result is expected to be a generic feature of any theory that 
obtains the superconductivity by doping the Mott insulator, and is not unique to the fractionalization route considered here. 
  
Remarkably, although the fractionalization {\it route} to
superconductivity is so very different from that in BCS theory,
the resulting superconducting phase itself is smoothly connected to the usual BCS superconductor\cite{Kiv,SNL,z2long}.  In particular, 
robust {\em universal} properties of the two superconducting phases will be the same. There will however be differences from a 
{\em mean field} description of a BCS superconductor. For instance, as we mentioned above, the superfluid density will be set by $x$
unlike in a mean field BCS superconductor. 

The usual BCS superconductor is 
understood as a condensate of charge $2e$ Cooper pairs while the superconductor obtained 
through fractionalization involves condensation of charge $e$ chargons. How can they 
be in the same phase? In particular, the flux quantization in the BCS superconductor
is in units of $hc/2e$ where the factor of $2$ in the denominator is supposedly a direct consequence 
of condensation of paired electrons. If the superconductor obtained by condensing the chargons is
in the same phase as the BCS one, it too must have $hc/2e$ flux quantization. This remarkable phenomenon
is made possible by the presence of the vison excitations when the electron is fractionalized.  
To see how this can come about, consider the effective $Z_2$ gauge theory Hamiltonian
describing the fractionalized phase Eqn. \ref{z2gh}. A finite doping requires 
inclusion of a chemical potential term 
\be
-\mu\sum_r N_r
\ee
that couples to the charge density. Now imagine leaving the fractionalized insulator
by condensing the chargons. 
It is instructive to focus on the regime
with large $K$, where a good description of the ground state
can be obtained by setting $\sigma_{ij} = 1$ on every link,
and taking the chargon field $b_r$ a space-time independent constant.
Consider placing an $hc/2e$ vortex at the (spatial) origin.
Upon encircling this $U(1)$ vortex at a large distance, the phase
of the chargon wavefunction must wind by $\pi$.  This is of
course not possible
with a smoothly varying phase field, but requires the introduction of a
``cut" running from the vortex to spatial infinity
across which the phase jumps by $\pi$.
The energy of this cut is, however, linear in its length
with a line tension proportional to $t_c |\langle b \rangle |^2$.
It thus appears that $hc/2e$ vortices are themselves confined,
and not allowed in the superconducting chargon condensate.
But imagine changing the sign of
all the $Z_2$ gauge fields, $\sigma_{ij}$,
which ``cross" the cut. 
This corresponds to placing a $Z_2$ vortex, {\em i.e}, a vison, at the origin.
These sign changes
``unfrustrate" the $XY$ couplings across the cut,
so that the line tension vanishes.  It is thus apparent
that a bound state of a vison and the $hc/2e$ $U(1)$ vortex (in the
phase of the chargon) can exist within the
chargon condensate.  It is this bound state which
corresponds to the elementary BCS vortex in the conventional
description of a superconductor.

It is worth emphasizing that both the ``naked" $hc/2e$ $U(1)$ vortex
and the vison are confined in the superconducting phase.
For example, the energy cost to pull apart {\it two} visons
also grows linearly with separation.  To see this,
introduce two visons by 
changing the sign of the $Z_2$ gauge field
along an interconnecting ``line".
Due to the chargon condensate which breaks the $Z_2$ gauge symmetry
making the gauge field ``massive",
{\it each} negative bond costs
an energy proportional to $t_c$, implying linear confinement.

Thus, the chargon condensate does have flux quantization in units of $hc/2e$. Note that we have only argued that
the energy cost of a $hc/2e$ vortex is {\em finite}. The question of whether this finite energy cost is nevertheless 
low enough that a pair of well-separated $hc/2e$ vortices are cheaper than a single $hc/e$ vortex is a different one, and will
be considered later. 

If chargon condensation is simply another viewpoint of a regular BCS superconductor, how are we to think of the BCS quasiparticles?
These are simply the spinons. Indeed, once the chargons are condensed, the visons get attached to $hc/2e$ flux as discussed above.
Consequently, in the absence of any vortices in the system, we may set the $Z_2$ gauge field to be one on all the bonds. The
Hamiltonian describing the spinons then becomes identical to the usual BCS Hamiltonian for {\em electrons} in a regular superconductor. 
In the presence of a $hc/2e$ vortex, the spinons see the $Z_2$ gauge flux associated with the vortex. Consequently, when a spinon is transported all
the way around a $hc/2e$ vortex, it picks up a phase change of $\pi$. This is exactly as required in a BCS superconductor if one works in a basis 
where the quasiparticles are made charge neutral (see Section \ref{vp}). 
It is indeed natural that the spinons (which are charge neutral by definition) become the 
BCS quasiparticles in this ``neutralized'' basis.

Thus, the chargon condensate does indeed correspond to a superconductor that is smoothly connected to a BCS superconductor.

\subsection{Pseudogap physics}
\label{pg}
The description of superconductivity given above is most natural in the underdoped samples where the approach of doping the Mott insulator 
may be considered reasonable. As explained in the
previous subsection, superconductivity is an inevitable low temperature consequence of the fractionalization if the doping is not too small. 
The fractionalization picture also unavoidably leads to the pseudogap physics that is observed in the underdoped materials. 

Consider the properties of 
an underdoped sample as a function of decreasing temperature.  A description of the properties of the system in terms of the fractionalized excitations
(the chargons and spinons) will become possible below a certain temperature scale. In the fractionalized insulator, this 
scale is essentially the vison gap, and 
is already non-zero (by assumption) in the undoped material. Upon doping, this scale will vary smoothly and would presumably decrease - with the 
additional assumption that Fermi liquid behavior is recovered in the heavily overdoped material, this scale would eventually go to zero with doping. 
An effective Hamiltonian
of the kind described in Eqn. \ref{z2gh} will then be a suitable description of the system at temperatures below this scale. 
Superconductivity develops at a 
much lower temperature scale which would go to zero as the doping is reduced. What are the properties of the system in the temperature window between
the superconducting transition and the temperature scale for fractionalization? To get some feeling for the physics in this regime, note that 
at temperatures well below the fractionalization scale (but above the superconducting $T_c$), the fluctuations of the $Z_2$ gauge field may be ignored so that the spinons and chargons have integrity. The spinon Hamiltonian is the same outside the superconductor as it is inside. Thus, the spin physics is expected not to vary much 
across the superconducting transition. More formally, since the $Z_2$ gauge fluctuations are suppressed, the spinon Hamiltonian 
explicitly describes a system with a spin pseudogap due to the presence of ``pairing'' terms. Thus there will be a suppression of the 
spin carrying excitations even above the superconducting transition temperature in the underdoped material. 

We emphasize that this pseudogap phenomenon is a rather general consequence of fractionalization as described here. Fractionalization liberates the
spin of the electron from it's charge. Consequently, the {\em number} $\sum_r f^{\dagger}_rf_r$ of the resulting spinon degrees does not need to be conserved by the dynamics. The Hamiltonian describing the spinons only needs to allow for spin conservation. Thus, singlet spinon ``pairing'' terms are generically 
allowed in the spinon Hamiltonian. This then gaps out the spinons.

When the electron is fractionalized, 
an electron added to the system will decay into
a spinon and chargon.  This has direct implications\cite{z2short,crtny} for
electron photoemission experiments.  Since the electron decays
one does not expect a sharp spectral feature in photoemission.
More formally, in this regime the electron propogator, $G(r,\tau)$, can be roughly expressed
as a product of the chargon and spinon propogators, $G_c$ and $G_s$:
\begin{equation}
\label{elgf}
G(r, \tau) \approx G_c(r, \tau) G_s(r, \tau)  .
\end{equation}
The spectral functions for the spinons and chargons ($A(k,\omega) = -\frac{1}{\pi}Im G(k,\omega)$)
will have sharp spectral features since these particles can propagate
coherently when the visons are gapped, but the {\it electron}
spectral function is a convolution of these two and will hence not
exhibit any sharp spectral features.  This is exactly as seen in the normal state ARPES spectra in the underdoped samples\cite{arpesud}. 
Now consider cooling the system into the superconducting state. As explained above, 
this requires condensation of the chargons so that
\be
G_c(r, \tau) \approx |< b >|^2  .
\ee
Then, from Eqn. \ref{elgf}, the electron Green's function just reduces to
\be
\label{elgfsc}
G(r, \tau) \approx |<b>|^2 G_f(r, \tau)  ,
\ee
and is simply 
proportional to the spinon Green's function inside the superconductor. 
Since the spinons propagate coherently, a sharp quasiparticle peak is expected - exactly as 
seen in the experiments\cite{arpesZ}. 
Moreover, 
since the {\it amplitude} of the peak is 
proportional to $| < b > |^2$,  it should become smaller as the 
superconductivity weakens, for instance, by reducing the doping. 
This is also borne out by the 
photoemission data\cite{arpesZ}. Thus, the qualitative trends in the underdoped
photoemission experiments can be 
well explained by assuming the electron decays into a chargon and a spinon. 

A further bonus of the fractionalization idea is the qualitative explanation of the strange 
electrical transport properties of the underdoped cuprates. This has been particularly emphasized by Anderson\cite{PWAbook}. 
The $c$-axis d.c. resistivity 
shows ``insulating" behavior increasing
rapidly upon cooling, whereas the in-plane resistivity is 
typically ``metallic"
and much smaller in magnitude.
Moreover, in a.c. transport a Drude peak is observed
in the $ab$ plane, but not along the $c-$axis.
This strangely anisotropic behavior,
difficult to understand within 
a conventional framework, follows naturally
if the fractions into which the electron decays 
reside primarily in the $ab$ plane. Transport along the $c$-axis requires hopping of {\it electrons} from layer to layer which is 
strongly suppressed
at low energies. 
Theoretically, by examining fractionalization in layered systems,
precisely such a phase where the fractions of the electron are deconfined within each layer but confined in the direction perpendicular to the layers
can be shown to exist\cite{topth} in gauge theory Hamiltonians of the form Eqn. \ref{z2gh}.

In the heavily overdoped non-superconducting samples, it is perhaps reasonable to expect Fermi liquid behaviour to emerge
at zero temperature. The ground state of the system then evolves from a fractionalized insulator through a conventional $d$-wave superconductor
to a Fermi liquid with increasing doping. The fractionalized insulator at zero or low doping also has various kinds of conventional order
(such as magnetism or charge ordering into, for instance, stripes). 

In the viewpoint adopted here, the driving force behind much of the physics is the 
fractionalization at low doping and the eventual evolution to the confined Fermi liquid at high doping. The details of how this evolution
from the fractionalized to the confined regimes actually occurs will presumably determine the properties around optimal doping. This is not
very well understood theoretically at present.

\subsection{Undoped insulator}
\label{ui}
The undoped Mott insulator is well-known to have Neel antiferromagnetism. The presence 
of such long range magnetic order has a number of immediate consequences for the 
{\em low energy} spin physics of the insulator. Indeed, there will be gapless spin wave excitations as a 
result of the broken spin rotation invariance. The physics of these spin waves and
their interactions will be well-described by an $O(3)$ non-linear sigma model field theory. Indeed, such a 
theory provides an excellent description\cite{CHN} of a number of experiments probing the low energy spin physics in the insulator.

But is the undoped Mott insulator really tame? A number of experiments\cite{arpesaf,iroa,raman,ntrnaf} have been performed over the last several years on the
insulator which perhaps hint that the magnetic order may be masking other important  physics that is much more fundamental
in determining the properties of the doped system. The most important among these are perhaps the ARPES results on the undoped cuprate\cite{arpesaf}
which do not exhibit a 
sharp quasiparticle peak 
at {\em any} momentum in the Brillouin zone. While it is indeed necessary to be cautious about photoemission experiments on 
insulators, this result is in striking contradiction with what is expected in a conventional Mott insulator even with magnetic
order. Early theoretical calculations\cite{KLR} of the motion of a single hole in a conventional antiferromagnet showed that there
is a finite non-zero quasiparticle residue of order $J/t$. Here $J$ is the magnetic exchange energy, and $t$ is the bare hopping 
amplitude of the holes. The commonly accepted estimate $t \approx 3J$ then gives a resonable weight to the quasiparticle
peak which is apparently not observed in the experiments. 

In view of this experiment, it is extremely interesting to contemplate the possibility that the magnetic ordering is actually coexisting with
the topological order and the associated fractionalization of the electron. As first pointed out by Balents et. al.\cite{NLII}, 
such a situation is indeed theoretically possible. Such a fractionalized antiferromagnetic Mott insulator will
have the same low energy spin wave physics as the more conventional antiferromagnet, but will nevertheless be in a distinct 
phase of matter. In particular, it will have gapped spinons, gapped chargons, and gapped visons as additional excitations. 
A precise {\em theoretical} distinction between the two phases may be obtained by asking for the ground state degeneracy
on topologically non-trivial manifolds\cite{topth}. 

The possibility of fractionalization in the undoped antiferromagnet is appealing for a variety of different reasons. The 
presence of long range magnetic order in the ground state implies that the long wavelength low energy physics 
will be determined by the spin waves just as in the conventional magnet. 
Thus the excellent description of the {\it low energy} spin physics by the $O(3)$ quantum non-linear sigma model 
is not sufficient to distinguish between the conventional and fractionalized antiferromagnets. Further, a 
fractionalized antiferromagnet offers a natural route to obtaining a photoemission spectrum with no sharp quasiparticle peak.
Yet further qualitative support 
is provided by mid-infrared optical absorption\cite{iroa} and Raman\cite{raman} 
measurements in the undoped material
which exhibit broad spectral features out to rather high energies,
not expected for the simple Heisenberg model. Finally, as we discussed earlier, fractionalization in the undoped system provides a 
simple rationale for the occurence of superconductivity in the doped system.

Despite all of these attractive features, the considerations above must be seen only as suggestive that the undoped insulator is fractionalized. 
Clearly, a direct detection of the fractionalization (or the lack thereof) will be extremely useful. A recent proposal\cite{acj} for an experiment
to achieve this suggests examining the ac Josephson effect in a device formed by sandwiching the undoped cuprate insulator between 
two superconductors. Observation of ac Josephson oscillations at a frequency $\frac{eV}{\hbar}$ as opposed to the usual $\frac{2eV}{\hbar}$
will firmly establish the presence of fractionalization in the insulator. For more details on this and other experiments to look for 
fractionalization in the undoped insulator, see Ref. \cite{acj}.

\subsection{Experimental tests}
\label{et}
The theory of the cuprates sketched above lends itself to a number of significant tests. 
It is central to the entire theory that the fractionalized phase have gapped vison excitations. This is true in all of the 
existing theoretical treatments of fractionalized phases (in $d > 1$) as we discussed at length in Section \ref{tatf}.
Furthermore, it is the presence of the visons that ensures that the superconductor obtained by condensing charge $e$ chargons is 
smoothly connected to a regular BCS superconductor with $hc/2e$ flux quantization. 

How may we detect the visons in experiments? The proposal of Ref. \cite{toexp,topth} is as follows. 
Consider a cylinder of highly underdoped superconducting cuprate with a hole drilled through it. Assume that this cylinder 
initially has a magnetic flux of $hc/2e$ trapped in the hole, but is in zero external field. Now imagine that one moves 
the system out of the superconducting phase either by heating or by other means. Then the electromagnetic flux will
escape out of the cylinder. On the other hand, if the non-superconducting state is fractionalized, the vison that is a 
part of the $hc/2e$ vortex will be unable to escape. Consequently, on moving back into the superconductor, 
this trapped vison will spontaneously nucleate a $hc/2e$ unit of electromagnetic flux. 

A number of other robust effects can be predicted for this kind of experiment. The most striking is to imagine doing the experiment
with a general integer multiple $n$ of $hc/2e$ flux that is initially trapped. A spontaneous final flux of $hc/2e$ will
appear only if $n$ is {\em odd}. This even-odd effect can be used to rule out other mundane explanations of the effect, if observed. 

As argued in Ref. \cite{toexp,topth}, the results of this experiment should be reasonably insensitive to all kinds of complications, such as disorder
or other broken symmetries such as magnetism or charge ordering. The most important caveat however is that the experiment needs to be done
at extremely low temperature. As the visons, if they exist, have a finite energy gap, they will always be 
thermally excited at any non-zero temperature. Consequently, the trapped vison will at any finite temperature eventually escape
in a thermal activation time $\tau_v \sim \tau_0 e^{\frac{E_{vison}}{k_B T}}$ where $E_{vison}$ is the vison gap, and $\tau_0$ is an attempt time. 
Thus, the experiment will need to be done in a time scale that is faster than this vison escape time. Neither the vison gap or the attempt time
can be reliably estimated theoretically. Ref. \cite{toexp,topth} suggested that the vison gap would be of roughly the same order as the spin gap 
in the underdoped samples. However, it is necessary to emphasize that a reliable calculation of the ratio of the vison gap to the spin gap
is not available at present. 

Another test of the fractionalization scenario for the underdoped cuprates
was pointed out a long time ago by Sachdev, Nagaosa, and Lee\cite{SNL}. They observed that
a superconductor that descends from a fractionalized insulator has regimes in which
the energy cost of an $hc/e$ vortex is smaller than two isolated $hc/2e$ vortices.
The basic physics underlying this observation is as follows. The vison gap is non-zero in the fractionalized 
insulator. On moving into the superconducting state, it is the presence of the vison that 
enables the existence of a finite energy $hc/2e$ vortex. The core energy of this vortex 
would then have to include the cost of having a  vison, and consequently, should be non-zero even at the transition
to the insulator. On the other hand, the core energy of a $hc/e$ vortex will go to zero. For both vortices,
the superflow energy will go to zero at the superconductor-insulator transition. Thus, close to the transition, 
the energy to create a single $hc/e$ vortex will be smaller than the energy to create a pair of well-separated 
$hc/2e$ ones. 
Thus observation of stable $hc/e$ vortices in the superconducting phase
would be an indirect test of the fractionalization in the ``normal" state. It is crucial 
to note that this argument relies on the sasumption that the superconductor-insulator transition is second order. 

Considerable caution is required in trying to observe these stable 
$hc/e$ vortices in experiments. The force between two $hc/2e$ vortices is
always repulsive at large separation (much bigger than the core size) where it is dominated
by the superflow. Thus it is necessary for two well-separated $hc/2e$ vortices to
overcome the superflow energy barrier and get close enough before the gain 
in core energy of the $hc/e$ vortex can provide for the attraction to bind them together. 
In practice, depending on the dynamics and the history of the sample, it may be possible
for $hc/2e$ vortices to be observable in some highly metastable state 
even in a regime in which a single $hc/e$ vortex has lower energy than a pair of 
$hc/2e$ ones.

\section{Discussion}
\label{disc}
The theory of the cuprates outlined here has, we believe, considerable phenomenological appeal. 
Apart from providing a simple route from the Mott insulator to the superconductor upon doping, it also helps understand
qualitatively several of the most puzzling phenomena in the ``normal'' state. We now 
briefly reemphasize some of these, and comment on some weak points. We also comment briefly on an alternate
theory that nevertheless shares many features with the present one.

Some features of the theory are generic to any picture of the superconductivity that remembers the proximity to the Mott insulator. 
An example is the result $\rho_s \propto x$.  

One of the major strengths of this theory is it's simple explanation of the ARPES results. In particular, 
it is very natural that the electron spectral function is very broad above the superconducting transition, but 
becomes sharply peaked below. While this is certainly evident in the ARPES results at the $(\pi, 0)$ point, the 
vicinity of the nodal points ({\em i.e} along $(\pi, \pi)$) has apparently not been examined carefully 
yet in experiments on underdoped cuprates\cite{Valla}.
 
The presence of the spin gap in the underdoped side is simple to understand in this theory. Indeed the spin physics is expected 
to not change significantly across the superconducting transition in the underdoped samples, as has been established in a 
number of experiments. Qualitatively, the anisotropic electrical transport - ``metallic'' in the plane and ``insulating''
perpendicular to it is also easily understood if the fractions of the electron are confined to the plane.

We have reconciled these spin-charge separation ideas with the occurence of antiferromagnetism at half-filling
by suggesting that the antiferromagnetic state at half-filling coexists with the fractionalization. This is indeed 
theoretically possible, and finds some support in experiments as was discussed in Section \ref{ui}. However, we have not provided a reason
for why the cuprates {\em prefer} to always order magnetically at half-filling. It is also not very clear in
this picture why the magnetism would be so easily destroyed upon doping. Chargon motion presumably does not frustrate the 
magnetic ordering in a fractionalized antiferromagnet, unlike hole motion in a confined magnet. 
 
Observation of the visons
would be a strong confirmation of the theory. On the other hand, if it can be established that the visons do not exist, or do so
with extremely small energy gaps, then it would be a severe constraint on the direct relevance of these fractionalization ideas
to the physics of the cuprates.

Very recently, the first results on the experiments discussed in Section \ref{et} have become available\cite{bonn,wynn,kirt}.
Ref. \cite{bonn,kirt} looked for the flux-trapping effect due to trapped visons in underdoped YBCO and BSCCO
respectively, and so far, have not seen any signatures of the vison.  
These negative results may be translated into a rough bound on the vison 
gap in the underdoped cuprates - at present, this bound is in the energy range $100 - 350 K$.  
The uncertainty is mainly due to
lack of a good estimate for the attempt time $\tau_0$. 
Ref. \cite{wynn} looked for, and did not find, stable $hc/e$ vortices in underdoped YBCO samples. 
We have already discussed some of the caveats on this test of the fractionalization idea. Ignoring these caveats, this negative result 
also leads to a rough bound of about $120 K$ on the vison gap. It is important to emphasize that while these results are
not very encouraging for the fractionalization scenario, they are not yet sufficient to rule it out due to the 
largeness of the bounds on the vison gap. Experiments that are currently underway\cite{kam} 
should be able to either place much more stringent bounds or 
observe the predicted effects.

We now briefly comment on an alternate theory of the cuprates which has some superficial similarity with the one discussed in this paper, but is 
distinct in some important respects. This is the $SU(2)$ gauge theory proposed by Lee, Wen, Nagaosa and coworkers\cite{su2}. 
In contrast to the theory discussed above, the $SU(2)$ gauge theory assumes that the system is always in some conventional 
(in the sense of no fractionalization) quantum phase at zero temperature. However, there still is some notion of ``spin-charge separation''
at short length/time scales in some temperature window that does not extend to the lowest temperatures. This scenario, if correct, 
would possibly retain some of the appealing features of truly fractionalized systems, without 
actually requiring the presence of such true fractionalization anywhere in the vicinity of the observed phase diagram. 
In particular, it has been 
argued that there will be no visons observed if such a scenario were correct. The actual nature and 
meaning of this short length scale
``spin-charge separation'' awaits clarification.

The starting point of the theory is a slave particle mean field
treatment of the $t-J$ model the results of which correspond quite closely to the observed phase diagram of the cuprates. 
Fluctuations about the mean field are however large, and are argued to be necessary to correctly describe the physics. The 
slave particle representation introduces an $SU(2)$ gauge symmetry - thus the theory of fluctuations about the mean field is a 
strongly interacting gauge theory. The pseudogap regime is associated with a ``breaking'' of the $SU(2)$ gauge
symmetry down to $U(1)$. It is suggested that the remaining $U(1)$ gauge degrees of freedom strongly scatter but do not confine the 
slave particles upto a temperature scale that goes to zero with the doping and hence is much smaller than $J$. At lower temperatures, there is 
confinement leading to conventional phases of matter.

One way to think about this proposed scenario is that the properties of the system are governed by the proximity to an 
{\em unstable} zero temperature non-Fermi liquid fixed point. Such a proximity would generate a wide window of 
anamolous behaviour at intermediate temperatures which finally gives way to conventional physics at the lowest temperatures. 
This is then reminiscent of (though perhaps distinct from) theories that suggest proximity to quantum critical points
between conventional phases\cite{CSY,so5,rbl} as the root of anamolous finite temperature behaviour in the cuprates. 

Several aspects of these ideas with short length scale ``spin-charge separation'' are 
very attractive for the cuprates, particularly if 
visons are not detected in experiments. Nevertheless their conceptual basis seems to require further elucidation.

\section{Conclusions}
\label{conc}
In this article, we have discussed the phenomenon of electron fractionalization in spatial dimensions $d > 1$. In the first 
part of the paper, we described the properties possessed by such a phase, should it exist. We argued that an effective theory of such a phase will
take the form of a $Z_2$ gauge theory. This conclusion is supported by a variety of different theoretical analyses of fractionalization, 
all of which lead to the same description. A precise theoretical characterization of such a 
phase may be obtained through the notion of topological order.

An important open issue is the demonstration of the presence of fractionalized phases in specific
microscopic models of interacting electrons. Some progress has been reported in Ref. \cite{lhull}.

In the second part, we considered the application of the fractionalization ideas to the cuprate materials. We discussed the attractiveness of 
these ideas for building theories of the cuprates, and how such ideas may be tested. An important theoretical question, should visons not be 
found, is whether it is possible to still retain the notion of fractionalization in any consistent way to understand the cuprates.

The perspective of this article was strongly influenced by my collaboration with Matthew P.A. Fisher whom I wish to thank
for innumerable discussions. I thank Courtney Lannert for a collaboration reported in Ref. \cite{crtny}. I am grateful also to 
P.A. Lee, R. Moessner, K.A. Moler, S. Sachdev, S. Sondhi, and X.G. Wen for much useful input.

\end{multicols}

\begin{references}
\bibitem{note1} In  general, integrating out high energy modes will also generate
other interactions not present in the original microscopic theory.
\end{references}
\end{document}